\documentclass[%
reprint,
superscriptaddress,
nobibnotes,
 amsmath,amssymb,
 aps,
 prl,
floatfix,
]{revtex4-1}

\usepackage{graphicx}
\usepackage{dcolumn}
\usepackage{bm}
\usepackage[dvipsnames,table]{xcolor} 

\usepackage{units}
\usepackage{lipsum}
\usepackage{xcolor}
\usepackage{siunitx} 		
\usepackage[symbol]{footmisc}
\usepackage{hyperref}

\begin{document}

\preprint{APS/123-QED}

\title{Low-Density Hydrodynamic Optical-Field-Ionized Plasma Channels Generated With An Axicon Lens}
\author{R. J. Shalloo}
\thanks{Now at The John Adams Institute for Accelerator Science, Blackett Laboratory, Imperial College London, London SW7 2AZ, United Kingdom}
\affiliation{John Adams Institute for Accelerator Science and Department of Physics, University of Oxford, Denys Wilkinson Building, Keble Road, Oxford OX1 3RH, United Kingdom}

\author{C. Arran}
\thanks{Now at York Plasma Institute, Department of Physics, University of York, York YO10 5DD, United Kingdom}
\affiliation{John Adams Institute for Accelerator Science and Department of Physics, University of Oxford, Denys Wilkinson Building, Keble Road, Oxford OX1 3RH, United Kingdom}

\author{A. Picksley}
\affiliation{John Adams Institute for Accelerator Science and Department of Physics, University of Oxford, Denys Wilkinson Building, Keble Road, Oxford OX1 3RH, United Kingdom}

\author{A. von Boetticher}
\affiliation{John Adams Institute for Accelerator Science and Department of Physics, University of Oxford, Denys Wilkinson Building, Keble Road, Oxford OX1 3RH, United Kingdom}

\author{L. Corner}
\affiliation{Cockcroft Institute for Accelerator Science and Technology, School of Engineering, The Quadrangle, University of Liverpool, Brownlow Hill, Liverpool L69 3GH, United Kingdom}

\author{J. Holloway}
\affiliation{John Adams Institute for Accelerator Science and Department of Physics, University of Oxford, Denys Wilkinson Building, Keble Road, Oxford OX1 3RH, United Kingdom}

\author{G. Hine}
\thanks{Now at Oak Ridge National Laboratory, Oak Ridge, Tennessee 37831, United States}
\affiliation{Institute for Research in Electronics and Applied Physics, University of Maryland, College Park, Maryland 20742, USA}

\author{J. Jonnerby}
\affiliation{John Adams Institute for Accelerator Science and Department of Physics, University of Oxford, Denys Wilkinson Building, Keble Road, Oxford OX1 3RH, United Kingdom}

\author{H. M. Milchberg}
\affiliation{Institute for Research in Electronics and Applied Physics, University of Maryland, College Park, Maryland 20742, USA}

\author{C. Thornton}
\affiliation{Central Laser Facility, STFC Rutherford Appleton Laboratory, OX11 0QX, UK}

\author{R. Walczak}
\affiliation{John Adams Institute for Accelerator Science and Department of Physics, University of Oxford, Denys Wilkinson Building, Keble Road, Oxford OX1 3RH, United Kingdom}

\author{S. M. Hooker}
\thanks{simon.hooker@physics.ox.ac.uk}
\affiliation{John Adams Institute for Accelerator Science and Department of Physics, University of Oxford, Denys Wilkinson Building, Keble Road, Oxford OX1 3RH, United Kingdom}

\date{\today}

\begin{abstract}
We demonstrate optical guiding of high-intensity laser pulses in long, low density hydrodynamic optical-field-ionized (HOFI) plasma channels. An axicon lens is used to generate HOFI plasma channels with on-axis electron densities as low as $n_e(0) = \SI{1.5e17}{cm^{-3}}$ and matched spot sizes in the range $\SI{20}{\micro\meter} \lesssim W_M \lesssim \SI{40}{\micro\meter}$. Control of these channel parameters via adjustment of the initial cell pressure and the delay after the arrival of the channel-forming pulse is demonstrated. For laser pulses with a peak axial intensity of $\SI{4e17}{W.cm^{-2}}$, highly reproducible, high-quality guiding over more than 14 Rayleigh ranges  is achieved at a pulse repetition rate of \SI{5}{Hz}, limited by the available channel-forming laser and vacuum pumping system. Plasma channels of this type would seem to be well suited to multi-GeV laser wakefield accelerators operating in the quasi-linear regime.

This article was published in \text{Physical Review Accelerators and Beams} \textbf{22}, 041302 on 10 April 2019. \href{https://doi.org/10.1103/PhysRevAccelBeams.22.041302}{DOI: 10.1103/PhysRevAccelBeams.22.041302}

\noindent \copyright 2019 American Physical Society.
\end{abstract}

\pacs{Valid PACS appear here}
\maketitle

Many applications of high-intensity laser-plasma interactions require the propagation of high-intensity laser pulses through plasmas which are orders of magnitude longer than the Rayleigh range. One example of particular current interest is the laser wakefield accelerator (LWFA) \cite{Tajima:1979}, in which a laser pulse with an intensity of order \SI{E18}{W.cm^{-2}} propagates though a plasma, driving a trailing density wave. The electric fields generated within this plasma wave are of the order of the wave-breaking field $E_0 = m_e \omega_p c /e$, where $\omega_p = (n_e e^2 / m_e \epsilon_0)^{1/2}$ and $n_e$ is the electron density \cite{Esarey:2009, Hooker:2013jk}. For plasma densities in the range $n_e = \SI{E17} - \SI{E18}{cm^{-3}}$, $E_0 \approx 30 -\SI{100}{GV.m^{-1}}$, which is several orders of magnitude higher than the fields generated in radio-frequency machines. 

Plasma accelerators can drive compact sources of femtosecond-duration \cite{Buck:2011dg, Lundh:2011b, Heigoldt:2015cd} radiation via betatron emission \cite{Kneip:2010,Cipiccia:2011}, undulator radiation \cite{Schlenvoigt:2008, Fuchs:2009}, and Thomson scattering \cite{Phuoc:2012vb, Powers:2013bx, Khrennikov:2015gx}, with many potential applications in ultrafast science. In the longer term they could provide a building block for future high-energy particle colliders \cite{Leemans2009}. 

For LWFAs operating in the quasilinear regime \cite{Esarey:2009}, the energy gain per stage varies as $W \propto E_0 L_\mathrm{acc} \propto 1/ n_e$, and the required length of the stage varies as $L_\mathrm{acc} \propto 1/ n_e^{3/2}$. Hence reaching higher energy gains requires the drive laser to propagate over longer lengths of lower density plasma. 

To date, the highest reported electron energy generated in a LWFA is \SI{7.8}{GeV}, which was achieved by guiding intense laser pulses through a 200-mm-long plasma channel with an axial electron density of \SI{2.7E17}{cm^{−3}} \cite{Gonsalves:2019ht}.

\begin{figure*}[tb]
\centering
\includegraphics[width=\textwidth]{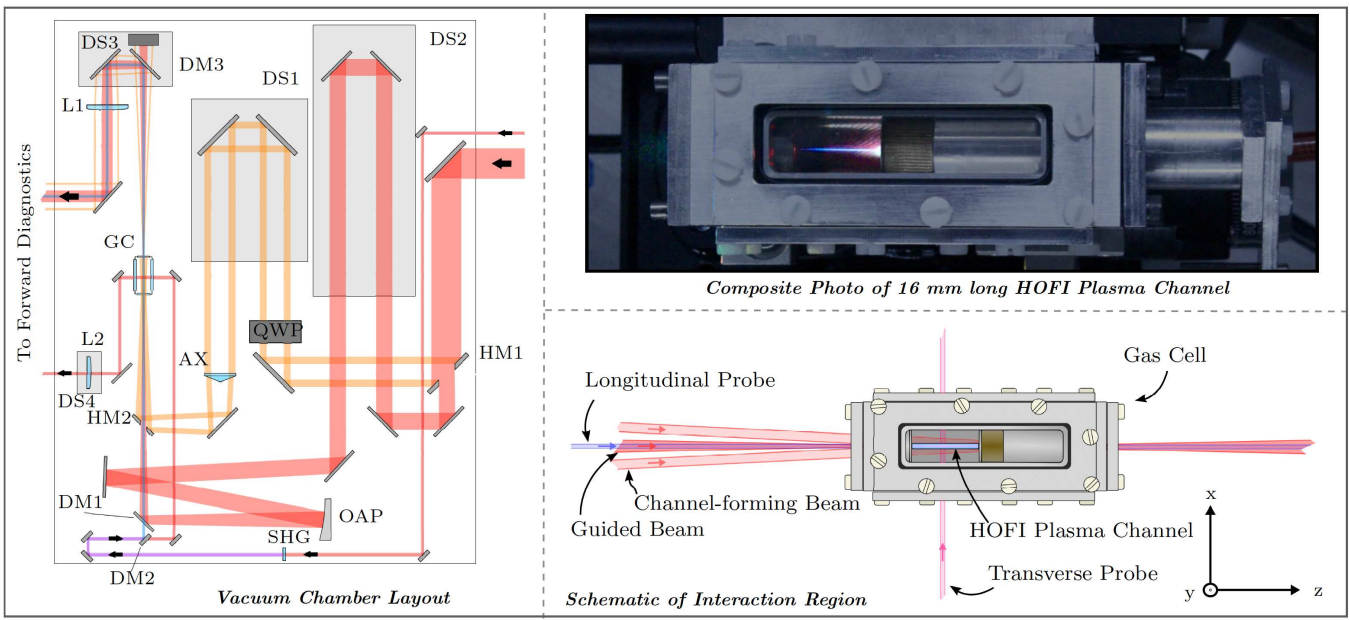}
\caption{\label{fig:experimentalLayout}\textbf{Left} Experimental layout: HM1 \& HM2 (mirrors with central holes), DM1 (\SI{810}{\nano\meter}:\SI{405}{\nano\meter} R:T dichroic), DM2 \& DM3 (\SI{405}{\nano\meter}:\SI{810}{\nano\meter} R:T dichroic mirrors), OAP (f/25 off-axis paraboloid), AX (refractive axicon), SHG (second harmonic generating crystal), QWP (quarter-wave plate), L1 (achromatic lens $f = 500$ \si{\milli\meter}), L2 (plano-convex lens $f = 250$ \si{\milli\meter}), GC (gas cell). \textbf{Top-Right:} Photograph of \SI{16}{\milli\meter} HOFI plasma channel. A composite of an image with chamber lights on and no plasma, together with an image of the plasma with the chamber lights off. \textbf{Bottom-Right:} Schematic of the interaction region showing the coupling of the four beamlines to the plasma source. }
\end{figure*}

Laser-plasma accelerators providing \SI{10}{GeV} energy gain per stage will require laser guiding through 100s of millimetres of plasma of electron density $n_e \approx \SI{E17}{cm^{-3}}$  \cite{Cros:2017, Leemans:2010zz}. Further, for many of the applications identified above it will be necessary to operate at pulse repetition rates, $f_\mathrm{rep}$, several orders of magnitude above the few hertz typical of today's GeV-scale LWFAs.

To date, the workhorse waveguide for LWFAs has been the capillary discharge waveguide \cite{Spence:2000fr, Butler:2002zza}. Capillary discharge waveguides have generated plasma channels up to \SI{150}{mm} long \cite{vanTilborgJ:2018fn}, with on-axis densities as low as approximately \SI{3E17}{cm^{-3}}. Electrical operation at $f_\mathrm{rep} = \SI{1}{kHz}$ has been reported \cite{Gonsalves:2016jc}, but high-intensity guiding at such high repetition rates has yet to be demonstrated. In recent work an additional nanosecond-duration laser pulse was used to heat and deepen the plasma channel through inverse bremsstrahlung heating \cite{Daniels2017}. Since in these devices the plasma channel is formed within a capillary structure,  their application to future, high-repetition-rate LWFAs will require the development of techniques to ensure that laser damage of the capillary discharge structure is avoided. 

We recently suggested \cite{Shalloo:2018fy} that hydrodynamic optical-field-ionized (HOFI) plasma channels could meet the challenges posed by future low-density, high-repetition-rate plasma accelerator stages. Since HOFI channels are isolated from any physical structure they are immune to laser damage and could therefore be operated at high repetition rates for extended periods.

In a HOFI channel a column of plasma is generated by optical field ionization; this expands hydrodynamically into the surrounding  gas, driving a radial shock wave, and forming a plasma channel along the axis of the plasma column. The concept is closely related to the hydrodynamic channels developed by Milchberg et al.\ \cite{Durfee:1993, Clark:1997we}; however, in a HOFI channel the axial density can be much lower since the plasma is heated by optical field ionization (OFI) rather than by laser-driven collisions. Our numerical simulations showed that an axicon lens could be used to generate  long plasma channels with on-axis densities of order of $n_e(0) \approx \unit[10^{17}]{cm^{-3}}$, matched spot sizes around $W_\mathrm{M}\approx \unit[40]{\mu m}$, and attenuation lengths (the length over which the intensity of a guided beam would drop by a factor of $e$) of order $L_\mathrm{att} \sim \SI{1}{m}$ \cite{Shalloo:2018fy}. Simulations show that only \SI{0.1}{mJ} of energy in the channel-forming laser is required per millimetre of HOFI channel, making realistic the generation of channels tens of centimetres long. The simulations were confirmed by experiments which showed the formation of short plasma channels generated at the focus of a \emph{spherical} lens. We note that Lemos et al.\ \cite{Lemos:2013gb, Lemos:2013ju} have generated short $n_e(0) \approx \SI{1E18}{cm^{-3}}$ HOFI channels produced by a spherical lens and have demonstrated guiding of low-intensity (\SI{1E15}{W.cm^{-2}}) laser pulses over 4 Rayleigh ranges \cite{Lemos:2018gx}.

In this Letter we demonstrate, for the first time, the formation of long, low density HOFI channels by an axicon lens. We use longitudinal and transverse interferometry to characterize the plasma dynamics and show the formation of plasma channels with axial electron densities as low as $n_e(0) = \SI{1.5e17}{cm^{-3}}$ and matched spot sizes in the range $\SI{20}{\micro\meter} \lesssim W_M \lesssim \SI{40}{\micro\meter}$. These channel parameters can be controlled by adjusting the initial cell pressure and the delay after the initial plasma formation. We demonstrate highly reproducible, high-quality guiding over more than 14 Rayleigh ranges of laser pulses with a peak intensity of \SI{4E17}{W.cm^{-2}} at a repetition rate of \SI{5}{Hz} (limited by the repetition rate of the laser system).

The experiments were performed with the Astra-Gemini TA2 Ti:sapphire laser at the Rutherford Appleton Laboratory, UK using the arrangement shown in Fig.\ \ref{fig:experimentalLayout} (see Supplemental Material for further details). For these experiments the laser provided linearly-polarized, \SI{430}{mJ}, \SI{45}{fs} pulses at a central wavelength of \SI{810}{nm}, in a beam of approximately \SI{55}{mm} diameter.

An annular beam --- the ``channel-forming'' beam, was apodized from the main beam by reflection from a mirror with a hole (HM1) and used to generate the HOFI channels. This beam was passed through a quarter-wave-plate (QWP), directed to a retro-reflecting delay stage (DS1), and then to a fused silica axicon lens of \SI{5.6}{\degree} base angle (corresponding to an approach angle of \SI{2.5}{\degree}). The beam focused by the axicon was directed into the target gas cell via a mirror (HM2) in which a hole had been drilled.

The beam transmitted through the hole in HM1 (the ``guided beam'') was directed to a second retro-reflecting delay stage (DS2); reflected from an off-axis paraboloid (OAP) of $f = \SI{750}{mm}$ focal length; and directed through HM2, and into the gas cell, by reflection from a dielectric mirror (DM1). The Rayleigh range of the guided beam was determined to be $z_\mathrm{R} = \SI{1.1(4)}{mm}$ from measurements of the longitudinal variation of the transverse size of the beam.

A separately compressed probe beam was used for longitudinal and transverse interferometry. A \SI{405}{nm} longitudinal probe beam was generated  by frequency doubling and directed into the gas cell by reflection from a dielectric mirror (DM2) and transmission through DM1. The remaining \SI{810}{nm} beam formed a transverse probe beam.

Hydrogen gas entered the variable-length gas cell (shown in Fig.\ref{fig:experimentalLayout}) via a \SI{4}{mm} inner diameter pipe, and left via a pair of coaxially-mounted stainless steel pinholes. The diameter of the front pinhole was $\SI{1500}{\mu m}$, to allow coupling-in of the axicon beam, and that of the rear pinhole was $\SI{750}{\mu m}$.

The beams leaving the gas cell longitudinally were reflected from a retro-reflecting delay stage (DS3) and collimated by an achromatic lens ($L1$) which acted as the primary lens of two Keplerian telescopes, one each for the \SI{810}{nm} and \SI{405}{nm} beams. The object plane of the telescopes, i.e.\ the front focal plane of L1, could be varied by adjusting DS3.

On leaving the vacuum chamber the beams were separated by a dichroic mirror. The \SI{405}{nm} and \SI{810}{nm} beams were  imaged by their respective secondary lenses and magnified onto a camera by  $5\times$ and $10\times$ microscope objectives, respectively. For the \SI{405}{nm} arm a folded-wave, Michelson interferometer was placed within the telescope.

The \SI{810}{nm} transverse probe beam was imaged onto a camera by a third Keplerian telescope, the primary lens of which (L2) was mounted on a stage (DS4) to allow  accurate imaging of the channel axis. An interferogram was formed by a folded-wave, Mach-Zehnder interferometer located within the telescope.

\begin{figure}[tb]
    \centering
    \includegraphics[width=\linewidth]{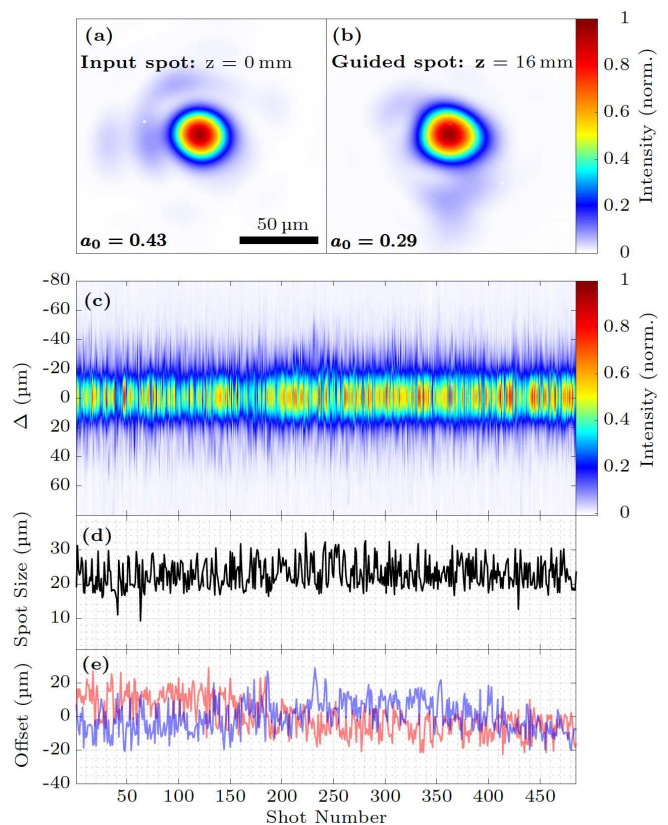}
    \caption{Transverse fluence profiles of the guided laser pulse at: (a) the entrance, and (b) the exit of a \SI{16}{mm} long HOFI channel, at $\tau = \SI{1.5}{ns}$. (c) shows, for each shot, the average of the horizontal and vertical fluence profiles through the centre of mass; the coordinate $\Delta$ has its origin at the centre of mass. In (d) the D4$\sigma$ spot size \cite{ISO11146-3} of the transmitted beam (averaged along the principal axes of the spot) is shown for each of the 485 shots; (e) shows the vertical (blue) and horizontal (red) offsets of the spot centre. In (a), (b), and (c) the peak fluence has been normalized to the highest value in each plot.}
    \label{Fig:Guiding_results}
\end{figure}

Figure \ref{Fig:Guiding_results} demonstrates guiding of pulses with a peak input axial intensity of $\SI{4e17}{W.cm^{-2}}$ in a \SI{16}{mm} long HOFI channel produced for an initial cell pressure of $P = \SI{60}{mbar}$ after a delay $\tau = \SI{1.5}{ns}$ between the arrival of the channel-forming pulse and the guided pulse. The transmitted transverse fluence profile of the guided beam measured at the channel exit (b) closely resembles that of the input beam (a), although it is very slightly elliptical. The orientation and degree of ellipticity was found to vary slightly with each shot, which is consistent with a shot-to-shot variation of the mode coupling arising from jitter in the transverse profile and pointing of the guided beam. To account for the slight ellipticity in the beam profile, the $D4\sigma$ spot size of the transmitted beam was calculated from the average of the spot size along the principal axes of the transmitted beam.  The on-axis density of the channel was found from the interferometric measurements described below to be $(3.6\pm 0.9) \times 10^{17}\, \mathrm{cm}^{-3}$.  In order to demonstrate the shot-to-shot stability of the HOFI channels, and to illustrate the potential for high repetition rate operation, the transverse intensity profiles of the guided beam were recorded for 485 consecutive shots, as summarized in Figs \ref{Fig:Guiding_results}(c) - (e). These data were recorded in $f_\mathrm{rep} = \SI{5}{Hz}$ bursts of approximately 12 shots every 40 seconds, with free flowing hydrogen gas during each burst. The repetition rate was that of the laser system, and the duration and cadence of the bursts were limited by the available vacuum pumps. The mean spot size of the transmitted beams was found to be $\SI{22.8 \pm 4.6}{\mu m}$, where the uncertainty is the root-mean-square (RMS) value over 485 shots. The RMS variation in the position of the centre of the transmitted beam was $\SI{14.1}{\mu m}$, which is approximately twice that of the input beam. The results shown in Fig.\ \ref{Fig:Guiding_results} demonstrate clearly the ability of HOFI channels to provide repeatable, high quality guiding of an intense beam over more than 14 Rayleigh ranges, with axial densities in the $10^{17}$ \si{cm^{-3}} range. The total energy transmission of the channel was found to lie in the 40-60 \% range. For the conditions for which the transverse intensity profile of the guided beam was most stable, an energy throughput of \SI{46(2)}{\percent} was recorded. Further work will be required to determine the relative roles of mode coupling and transmission losses in determining the total energy transmission.

The HOFI channels were characterized by longitudinal and transverse interferometry in short (\SIrange{2}{4}{\milli\meter}) and long (\SI{16}{\milli\meter}) cells respectively. Plasma channels were observed with both longitudinal and transverse interferometry, and the transverse phase profiles obtained were found to be consistent. For the longitudinal measurements, the plumes of gas emerging from the cell pinholes cause an unknown longitudinal variation of the plasma density; this introduces some uncertainty in the deduced axial electron density, although the shape of the transverse electron density profile is expected to be representative of the channel profile. Quantitative measurements of the electron density profiles of the channels were therefore obtained from the transverse measurements, as described below.

Both the transverse and longitudinal interferometry showed that the plasma channels had a slight azimuthal asymmetry (see also the Supplementary Information), which we attribute to asymmetries observed in the near-field profile of the channel-forming beam.  This asymmetry did not hinder channel formation --- as evident from the observation of high quality guiding ---  but it did mean that the transverse electron density profile could not reliably be deduced from the transverse interferometric data by the usual Abel inversion procedure.

The transverse data was analyzed as follows (see Supplemental Material for further details). It was assumed that the transverse electron density profile $n_e(x,y)$ could be written as a superposition of a finite number of basis functions. For each trial superposition, integration along the $x$-axis yielded the expected phase $\phi_\mathrm{trial}(y)$ imprinted on a transverse probe beam; this was fitted to the measured transverse phase $\phi_\mathrm{meas}(y)$ to yield $n_e(x,y)$. We used a basis set comprising the square-modulus of Laguerre-Gaussian functions since this can generate both guiding and non-guiding transverse profiles and ensure that the retrieved electron density is positive. Asymmetry was introduced by allowing the coordinates of centre position of the modes to be complex \cite{Kovalev:2016}. Tests with simulated data showed that this procedure could reliably reconstruct the transverse electron density profiles of plasma columns with both guiding and non-guiding transverse profiles (see Supplementary Information).  Good agreement with $\phi_\mathrm{meas}(y)$, with a reduced chi-squared of order unity, was typically achieved for a superposition of modes with radial index $m = 0$ and azimuthal indices $n \leq 3$.

\begin{figure}[tb]
    \centering
    \includegraphics[width=\linewidth]{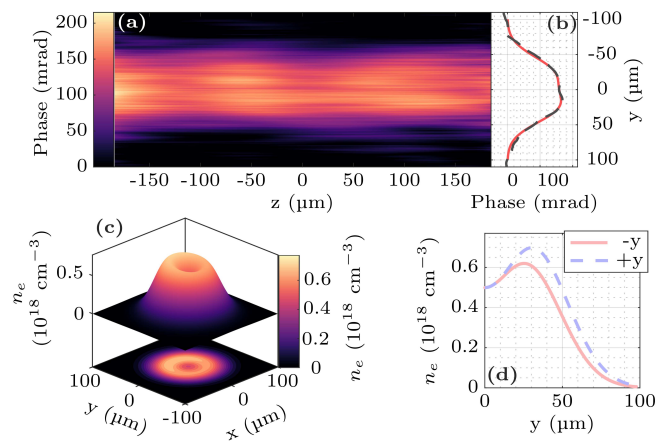}
    \caption{Transverse interferometric measurements of HOFI channels for an initial cell pressure of $P = \SI{80}{\milli\bar}$ and delay $\tau = \SI{1.5}{\nano\second}$. (a) shows the measured transverse phase map of the channel while (b) shows the longitudinally averaged phase (dashed gray) and a lineout from the transverse phase fitting procedure described in the text (solid red). The lower panel shows, for the retrieved electron density profile: (c)  a surface and contour plot of the retrieved electron density; and (d) lineouts along the positive and negative portions of the y-axis.}
    \label{Fig:PlasmaChannels}
\end{figure}

\begin{figure}[tb]
    \centering
    \includegraphics[width=\linewidth]{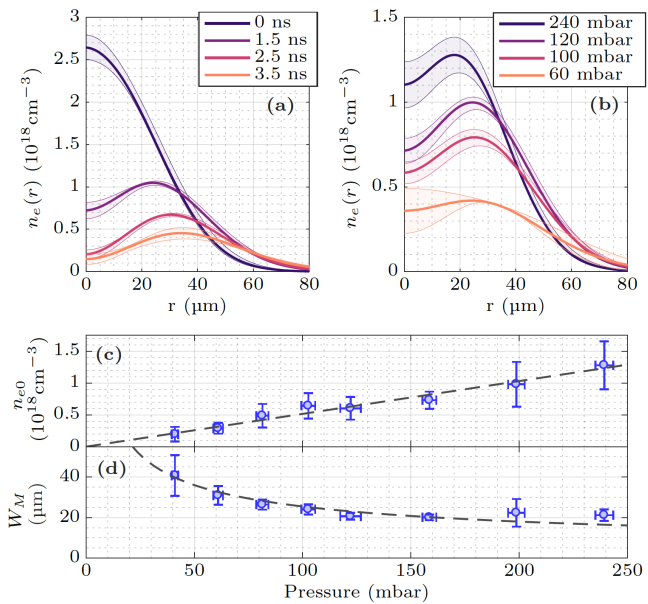}
    \caption{Variation of the rotationally-averaged radial electron density profile $n_e(r)$ of the plasma channel with: (a) delay $\tau$, for $P = \SI{120}{mbar}$; and (b) with initial cell pressure $P$, for $\tau = \SI{1.5}{ns}$. In  (a) and(b) the solid line shows the mean value, and the shaded region the standard error, obtained from averaging typically 5 shots. Also shown are the variation with $P$, for $\tau = \SI{1.5}{ns}$ of: (c) the on-axis density $n_e(0)$, together with a linear fit; and (d) the matched spot size $W_M$, together with a fit to $W_M \propto P^{-0.5}$.}
   \label{Fig:interferometryResults}
\end{figure}

Figure \ref{Fig:PlasmaChannels} shows an example of the analysis of the transverse interferometry data shown in (a). The measured phase averaged along the channel axis $z$, $\phi_\mathrm{meas}(y)$, is shown in (b), together with the fit to this lineout obtained from the analysis. It can be seen that the measured and fitted transverse phase profiles are in very close agreement. The retrieved electron density profile $n_e(x,y)$ is plotted in Fig.\ \ref{Fig:PlasmaChannels} (c) and (d), showing clearly the formation of a channel, with a slight azimuthal asymmetry. This asymmetry was found to be small; for the data shown in Figs \ref{Fig:PlasmaChannels} and  \ref{Fig:interferometryResults} the fractional azimuthal variation in density at $r = W_M$ was between 4\% and 11\%.

Figure \ref{Fig:interferometryResults} (a) presents the temporal evolution of the rotationally-averaged, retrieved electron density profiles for $P =  \SI{120}{\milli\bar}$, demonstrating the formation of a plasma channel over a few nanoseconds. The on-axis density and matched spot size of the plasma channel can be controlled by adjusting the cell pressure, as shown in Fig.\ \ref{Fig:interferometryResults}(b) -- (d). The matched spot size was calculated from a parabolic fit to the channel profile between the origin and the peak density of the shock; this procedure was found to be in good agreement with calculations of the matched spot size of the lowest-order mode calculated by a Helmholtz code \cite{Clark:2000dk}. It can be seen that in these experiments plasma channels with on-axis densities as low as $\SI{1.5e17}{\per\centi\meter\cubed}$ were observed, with matched spot sizes of approximately $\SI{40}{\mu m}$. As previously observed for HOFI channels generated with a spherical lens \cite{Shalloo:2018fy}, for a fixed delay $\tau$ the matched spot size decreases with cell pressure. The on-axis density is measured to be approximately proportional to the initial cell pressure $P$, and hence the plasma wavelength on-axis varies as $\lambda_p \propto P^{-0.5}$. A fit to the matched spot size then shows that under these conditions $W_M \approx (0.55 \pm 0.05) \lambda_p$. This scaling of the matched spot size with density is ideal for LWFAs operating in the quasi-linear regime, which require $\frac{1}{2}\lambda_p \lesssim W_M \lesssim \lambda_p$ to avoid relativistic self-focusing and reduced group velocity \cite{Leemans:2010zz}.

In summary we used an axicon lens to generate long (\SI{16}{mm}), low density ($n_e(0) < \SI{1E18}{cm^{-3}}$) hydrodynamic optical-field-ionized plasma channels for the first time. Plasma channels were generated with axial densities as low as $n_e(0) = \SI{1.5e17}{cm^{-3}}$ and matched spot sizes in the range $\SI{20}{\mu m} \lesssim W_M \lesssim \SI{40}{\mu m}$; these channel parameters could be controlled by adjusting the initial cell pressure and the delay $\tau$ after the arrival of the channel-forming pulse.

Highly reproducible, high-quality guiding of laser pulses with a peak axial intensity of $\SI{4e17}{W.cm^{-2}}$ was demonstrated over more than 14 Rayleigh ranges at a pulse repetition rate of \SI{5}{Hz}, limited by the laser system.

It is worth noting that in this work only \SI{0.5} - \SI{0.8}{mJ} of channel-forming laser energy was required per millimetre of channel, more than an order of magnitude less than for channels formed with a spherical lens, and within a factor of 5 of that predicted by simulations \cite{Shalloo:2018fy}. We note that with a suitable gas cell or long gas jet the axicon used in this work could generate plasma channels up to \SI{570}{\milli\meter} long.

The work described here confirms the potential for HOFI channels to operate as free-standing plasma channels with axial densities in the $\SI{e17}{cm^{-3}}$ range, with matched spot sizes ideally suited to LWFAs operating in the quasi-linear regime, with lengths of hundreds of millimetres, and with the possibility of operation at high repetition rates for extended periods \cite{Shalloo:2018fy}. Plasma channels with these characteristics would be an ideal medium in which to drive high-repetition-rate, multi-GeV energy gain, metre-scale plasma stages suitable for compact light sources and possible future TeV colliders \cite{Schroeder2010}.

This work was supported by the UK Science and Technology Facilities Council (STFC UK) [grant numbers ST/J002011/1, ST/P002048/1, ST/M50371X/1, ST/N504233/1, ST/R505006/1]; the Engineering and Physical Sciences Research Council [Studentship No.\ EP/N509711/1]; the Helmholtz Association of German Research Centres [Grant number VH- VI-503]. This material is based upon work supported by the Air Force Office of Scientific Research under award numbers FA8655-13-1-2141 and FA9550-18-1-7005. This work was supported by the European Union's Horizon 2020 research and innovation programme under grant agreement No. 653782. GH and HMM were supported by US Dept. of Energy (Grant No. DESC0015516), Dept. of Homeland Security (Grant No. 2016DN077ARI104), National Science Foundation (Grant No. PHY1619582), AFOSR (Grants No. FA9550-16- 10121 and FA9550-16-10284)


\bibliography{References}

\end{document}